\documentclass[aps,prl,preprint,showpacs,floatfix]{revtex4}
\usepackage{epsfig}
\usepackage{psfrag}
\begin{document}
\title{Morphological Phases of Crumpled Wire}

\author{ N. \ Stoop, F.\ K. \ Wittel, and H.\ J.\ Herrmann} 
\affiliation{Computational Physics for Engineering Materials, ETH Zurich, Schafmattstr. 6, HIF, CH-8093 Zurich, Switzerland} 
\date{\today}
   
\begin{abstract}
We find that in two dimensions wires can crumple into different morphologies and present the associated morphological phase diagram. Our results are based on experiments with different metallic wires and confirmed by numerical simulations using a discrete element model. We show that during crumpling, the number of loops increases according to a power-law with different exponents in each morphology. Furthermore, we observe a power-law divergence of the structure's bulk stiffness similar to what is observed in forced crumpling of membranes.
\end{abstract}
 

\maketitle

Crumpling is omnipresent in nature, occurring on all length scales, ranging from blood cells that crumple in order to pass through capillaries up to the formation of the Swiss Alps. The crumpling of spatially extended membranes became a challenging research topic of strong interest over the last decades. Thorough research, both experimental, theoretical and numerical was conducted to discover basic statistical properties \cite{gomes1,nelson1} like the scaling of strength and energy \cite{lobkovsky1, witten2, gompper} and landscapes in phase space \cite{kramer,blair} of crumpled structures. Surprisingly the question of packing of wires or polymer chains in two dimensions has attracted far less attention until recently \cite{gomes,boue-etal}.

In this letter, we present the first morphological phase diagram of crumpled wires in two dimensions. Metal wires with different material behavior are crumpled by injecting them into a circular cavity from opposing sides. To study the material dependence of the morphology, we keep the general geometric setup fixed, instead of allowing e.g. for the injection angle or cavity shape to be varied like in \cite{gomes}. Our experiments show that plastic yielding and friction are the essential parameters determining the morphology. We find excellent agreement with computer simulations and construct a full morphological phase diagram. We further analyze the morphologies by means of the number of loops as function of the packing density, which exhibits a power-law behavior with different exponents for each morphology. Finally we show that the stiffness of the crumpled structures follows a power law increase in the packing fraction, similar to observations on membranes \cite{witten2,astrom1}. The stiffness, however, does not fully diverge due to the finite wire thickness, resulting in jammed states where percolating force chains are formed by contacting wire segments.
\begin{figure*}[htb]
  \begin{center}
    \epsfig{bbllx=17,bblly=455,bburx=560,bbury=790,file=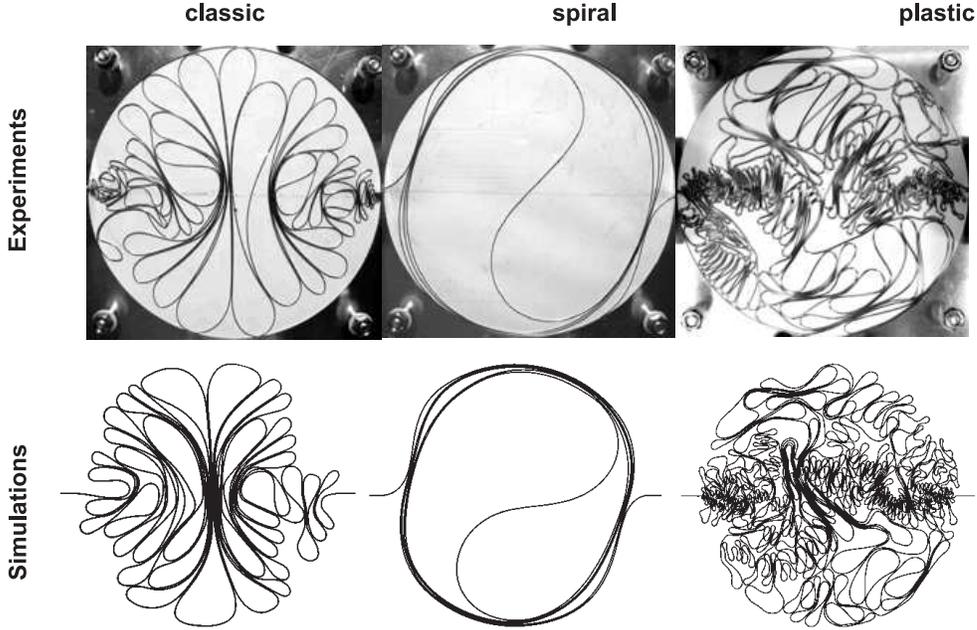,  width=14.5cm}
    \caption{\label{fig:1} \small Crumpled wires in a circular cavity. Experiments (top row) and simulations (bottom row), showing the three different morphologies. Materials are steel wire (1.5125) $d$=0.8mm without and with silicon oil (left) and (middle) and brazing solder (Sn99Cu1) $d$=1mm (right). Simulation parameters are $f$=100, $\kappa_\theta$=0.04;$\mu_{st}$=5;$s$=0.08 for the classic (left), $\kappa_\theta$=$\pi$/2;$\mu_{st}$=0;$s$=0.04 for the spiral (middle) and $\kappa_\theta$=0.02;$\mu_{st}$=5;$s$=0.04 for the plastic (right) morphology. Simulation snapshots are shown for similar packing densities as in the respective experiments.}
  \end{center}
\end{figure*}

In our experiments a wire is pushed from opposing sides into a metallic cylindrical cavity of radius $r$=10cm having a width of one wire diameter $d$ with a transparent acrylic glass top to allow for observation. The setup is based on experiments by Gomes \cite{gomes}, with the crucial difference, that we are pushing the wire in a controlled way by two sets of counter rotating rollers driven by large forces. With this setup we crumple wires of steel ($d$=0.8mm) and brazing solder ($d$=1mm) - materials with differing mechanical properties.

We model the wire by point masses, connected by tensile springs. Bending stiffness is considered by rotational springs attached to each node. We choose Hook's law
\begin{equation}
F_i = - k_1 (\lambda_0 - \lambda_i)\quad\text{and} \quad M_i = -k_2 \zeta_i \label{bendingmoment}
\end{equation}
with the rest length $\lambda_0$, the angle $\zeta_i$ at node $i$ and the two proportionality constants $k_1$ and $k_2$ for the tensile forces and bending moments respectively. From the continuum limit \cite{domokos} it follows that $k_1=EA/\lambda_0$ and $k_2=EI/\lambda_0$ with Young's modulus $E$, wire cross-section $A$ and second moment of inertia $I$. For the numerical results we fixed $E$=1275, corresponding to $k_1$=1000 and $k_2$=60 for a circular wire with diameter $d$=1 and $\lambda_0$=1. We use the dimensionless ratio $f=r/d$ of cavity radius to wire diameter to specify the effective system size.

We describe plastic deformation of wires in the rotational springs, using a simple linear flow rule with yield threshold moment of $k_2\alpha_\theta$ and slope $sk_2$ (0$\le$$s\le$1). Therefore $M_i$ in Eq.\ref{bendingmoment} holds for $\zeta \le \alpha_\theta$, while for  $\zeta > \alpha_\theta$ $M=-k_2\alpha_\theta-(\zeta -\alpha_\theta)sk_2$. Unloading is always along the elastic path with slope $k_2$. This is a linear approximation to the bending stress-strain relation, as can be found for example in Refs. \cite{dobrovolskii}. In the following, we will use the discretization-independent yield curvature $\kappa_\theta = \alpha_\theta / \lambda_0$ as plasticity parameter. 

For a realistic simulation considering friction at wire-wire and wall-wire contacts proved to be crucial. We therefore implemented a simple Coulomb's friction law $F_{st}\le\mu_{st}F_N$ with friction coefficient $\mu_{st}$ and force normal to the contacts $F_N$. Above $\mu_{st}F_N$, dynamic friction sets in with $F_d=\mu_d F_N$, opposed to the relative tangential movement of contacts. Below a threshold velocity $v_{th}$, static friction sets in again \cite{margolis}. For numerical reasons, we add a small viscous damping on the translational and rotational degrees of freedom.

We follow the time evolution of the system by integrating the equations of motion of all nodes. We choose $f$=100 to match the experimental setup. During the simulation we push new elements into the cavity, starting from an initially cavity-spanning straight wire. The simulation stops, when a sudden increase of the injection force appears. At this point, we consider the system as jammed. The experiments end, when the resistance of the crumpled structure becomes so large, that the rollers are not able to push more wire into the cavity. 

The crumpling starts with an initial buckling of the cavity-spanning wire in the up- or downwards direction. The symmetry of the wire is broken by small deviations from the ideal positions. This solution is stable until the wire contacts the cavity wall for the first time. Depending on yield threshold $\kappa_\theta$ and static wire friction coefficient $\mu_{st}$, three different morphologies can be distinguished: 

(1) With high cavity friction $\mu_{st}$ a  morphology emerges that we call {\it classic}, since it corresponds to the observation of Ref. \cite{gomes}. As can be seen in Fig.\ref{fig:1}(left), the wire immediately forms cascades of loops of decreasing sizes. Since wire is inserted on both sides, strong symmetry is present. In simulations of purely elastic wires, this symmetry is preserved until the structure is jammed, while experimentally, it is broken when plastic flow sets in for high packing densities. Cascading loops and strong symmetry are the two criteria for this classical morphology.

(2) Elastic wires with low $\mu_{st}$ form a {\it spiral} pattern, with the turning direction chosen spontaneously (see Fig.\ref{fig:1}(middle)) comparable to Ref.\cite{boue-etal}. The spiral winds up more when more wire is inserted. Simulations reveal that this pattern is maintained for purely elastic wires until the entire cavity is filled. In the experiment, however, a critical packing density might exist where plastic deformations set in. In that case, loop cascades appear on the left and right side of the spiral, i.e. a mixing of the classic and the spiral morphology can be observed.

(3) Typical for the {\it plastic} morphology is the loss of symmetry. This third morphology arises for highly plastic wires (Fig.\ref{fig:1}(right)). It starts similarly as the classic one by forming cascades of loops. In contrast to the classical morphology the axial symmetry is soon broken and large rearrangements of existing structures occur. Such reordering is observed on length scales ranging from a single loop up to the rearrangements of entire cascades. As a consequence, higher packing densities can be obtained than for the other morphologies. Note that the plastic phase is defined by the disorder and not by the material. Plastic and classic morphologies are separated by comparing the spatial distribution of curvature, which is concentrated near the cavity border for the classic phase (cf. the location of loops in Fig. \ref{fig:1}) \footnote{The cavity is divided into two regions of equal area by placing a circle at its center. If $A/B$$<$0.5, where $A$ ($B$) is the total squared curvature in the circle (surrounding ring), we consider the morphology as classic.}.
 
\begin{figure}
  \begin{center}
    \epsfig{bbllx=0,bblly=0,bburx=400,bbury=245,file=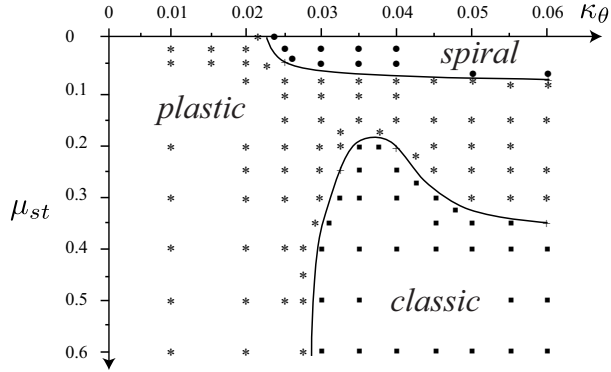, width=8cm}
    \caption{\label{fig:2}\small Morphological phase diagram of the spiral/plastic/classic morphologies obtained numerically for $f$=50, where the control parameters are the static wire friction $\mu_{st}$ and the elastic yield curvature $\kappa_\theta$. Each point is averaged over 9 realizations.}
  \end{center}
\end{figure}

Experimentally, we produced the classic, spiral and plastic morphologies using steel wire without and with silicon oil or brazing solder, each one represented by one point in the phase space. We investigated this phase space numerically in detail by changing yield curvature $\kappa_\theta$ from $0$ to $0.06$ and friction $\mu_{st}$ from $0$ to $0.6$, resulting in the morphological phase diagram of Fig.\ref{fig:2}. Simulations are limited to $f$=50, to be able to present a precise phase diagram in reasonable CPU-time, however preliminary results for $f$=100 do not show significant differences. For $\kappa_\theta=0$ we allow plastic deformation right from the start, while $\kappa_\theta=\pi/\lambda_0$ represents the elastic case. For $\mu_{st}$$>$0.6 and $\kappa_\theta$$>$0.06 we observed no further influence on the morphology. For all simulations $k_1$=1000, $k_2$=60 and $s$=0.05 were kept constant. We find that a direct transition from the spiral to the classic phase is not possible and a reentrant phenomenon is observed at 0.2$<$$\mu_{st}$$<$0.35 with increasing $\kappa_\theta$. For an explanation, consider the two requirements for building the symmetric structures of the classic phase: First, friction stabilizes the structures during packing. With $\mu_{st}$ between 0.08 and 0.18, this constraint is not sufficient and the loops rotate and rearrange. Consequently, the plastic phase emerges between the spiral and classic one. However, also plastic deformation has a stabilizing effect by dissipating stored energy, limiting the system's capability to rearrange. To find the classic phase at 0.2$<$$\mu_{st}$$<$0.35, small $\kappa_\theta$ is required to prevent rearrangements, along with friction. For too small $\kappa_\theta$, no cavity spanning loops form and the classic phase is unreachable. Consequently, the classic phase requires a minimal $\kappa_\theta$$\approx$0.027, leading to the observed reentrant phenomenon.

\begin{figure}
  \begin{center}
    \epsfig{bbllx=17,bblly=12,bburx=244,bbury=195,file=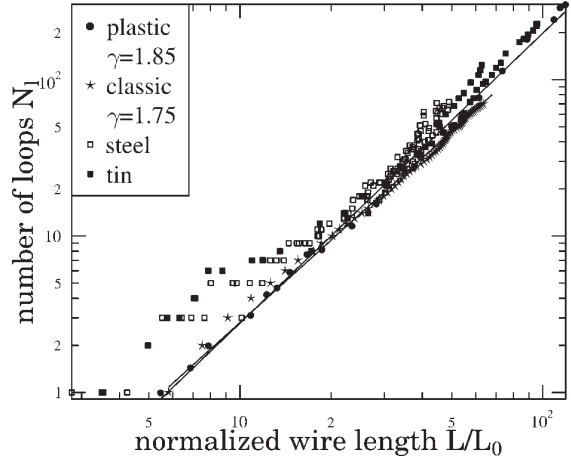, width=7.5cm}
    \caption{\label{fig:3}\small Scaling of the number of loops for the classic ($f$=50) and plastic ($f$=100) morphologies. To compare different systems sizes, $L/L_0$ was used as packing ratio (total inserted wire length $L$ and initial one $L_0$=$2R$). The number of loops for both morphologies follows the power law $N_l\approx \phi^\gamma$.}
  \end{center}
\end{figure}

The morphology of the crumpled wires can be quantified by the number of loops, the distribution of loop sizes and contact points, or the fractal dimension of the structure. In this letter we focus on the scaling of the number of loops $N_l$ as a function of the dimensionless packing density $\phi=d \cdot L/(\pi r^2)$, where $L$ is the inserted wire length. A loop is defined as an area that is surrounded by a wire segment with only one inner contact of inward wire surfaces. In experiments, the number of loops was counted and the total wire length was obtained by standard image analysis of digital images that were taken during the experiment. Gomes et al.\cite{gomes} report a pronounced shoulder for $\phi$$<$0.032, a power law asymptotic dependence $N_l \sim \phi^\gamma$ with $\gamma$=1.8$\pm$0.2 and a maximum packing density of $\phi$=0.14$\pm$0.006. The results of our simulations and experiments are given in Fig.\ref{fig:3}. We find exponent $\gamma$ to vary slightly for different morphologies. For the classical morphology we measure $\gamma$=1.75$\pm$0.03, for the plastic morphology $\gamma$=1.85$\pm$0.04 and for the spiral $\gamma$=0 up to $\phi$=1, as only two loops are present. Also a universal maximum packing density could not be observed.

\begin{figure}
  \begin{center}
    \epsfig{bbllx=121,bblly=515,bburx=355,bbury=715,file=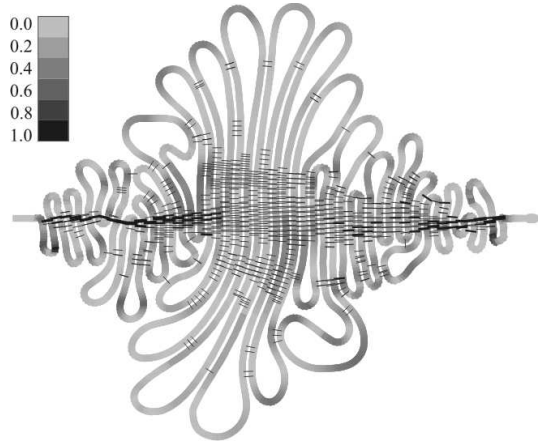, width=7cm}
    \caption{\label{fig:4}\small The jammed classic morphology. Force chains formed by contacts are clearly visible. Contact forces are represented by the thickness of the line segments, while the grey scale of the wire backbone represents its local elastic bending energy. Since this system includes plastic deformations, highest curvature does not necessarily imply highest bending energy.}
  \end{center}
\end{figure}

For a macroscopic analysis of the packing process, we measured the stiffness of the crumpled structures as function of the packing density $\phi$ via the force acting on the nodes in the insertion channels. As before, the simulation is stopped when the system is jammed, i.e. when straight parallel force lines from contacting wire segments exist between the injection channels (see Fig.\ref{fig:4}).

\begin{figure}
  \begin{center}
    \epsfig{bbllx=14,bblly=8,bburx=255,bbury=195,file=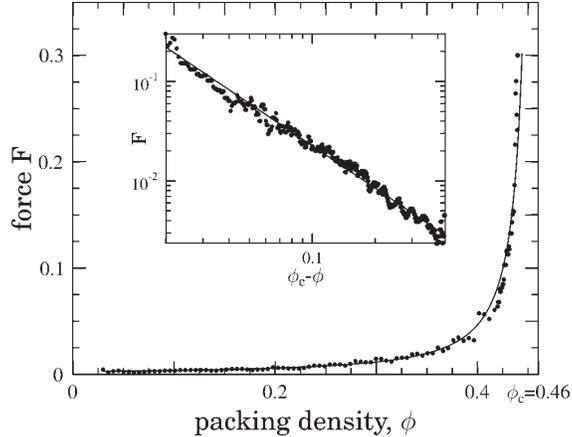, width=7.5cm}
    \caption{\label{fig:5}\small The insertion force $F$ as function of $\phi$. Simulation parameters are $k_1$=1000, $k_2$=60, $f$=25, $\kappa_\theta$=0.08, $\mu_{st}$=5, $s$=0.05. The best fit for the divergence is a power law $F\propto (\phi_c - \phi)^\beta$ with $\beta$=-1.43 (solid line).}
  \end{center}
\end{figure}

For computational reasons, these measurements were performed for different system sizes up to $f$=35. Parameters were chosen as $\kappa_\theta$=0.08, $s$=0.06 and $\mu_{st}$=5, corresponding to the classical phase. We find the force to scale as a power-law of the form
\begin{equation}
F \propto (\phi_c - \phi)^\beta
\end{equation}
with exponent $\beta$=-1.43$\pm$0.02. In Fig.\ref{fig:5}, $\phi_c$=0.46$\pm$0.01 which is substantially smaller than the theoretical limit $\phi_c$=1. For the purely elastic case of the classic morphology, we find $\beta$=-2.05$\pm$0.02 and $\phi_c$=0.54$\pm$0.02 (data not shown). Note that simulations for the spiral phase exhibit $\phi_c \approx 1$ with a deviant force scaling relation, in excellent agreement with the analytical results \cite{boue-etal}.

In this letter we presented three morphologies into which wires crumple inside a circular cavity. Using a discrete element model that incorporates plastic deformation and static friction, we showed that plasticity and friction are the two essential parameters determining the morphology. Three different experimentally found morphologies could be reproduced {\it in silico} by matching wire materials and friction, and the associated phase diagram was constructed. We showed the existence of a reentrant phenomenon due to two mechanisms for the stabilization of structures, friction and plastic deformation. The crumpled structures were analyzed by counting the number of loops as function of the packing density, showing a power-law behavior for experiments and simulations. The insertion force, which is difficult to obtain in experiments, was determined numerically and exhibits a power-law divergence with critical packing fractions substantially lower than 1. The same power-law divergence (with $\phi_c$$\approx$0.75, $\beta$$\approx$-1.85 \cite{astrom1,matan}) was also found for the forced crumpling of membranes in three dimensions, although the divergence in membranes comes from kinks \cite{witten2,witten1}. In wires, on the other hand we have no kinks and it is most probably a consequence of the decreasing size of the newly generated loops.

The presented work demonstrated the crucial role of plasticity and friction in crumpling processes, with its characteristic morphology. It is natural to generalize this work to the forced crumpling of membranes or the packing of DNA into viral capsids \cite{dnajungs,gomes2}. Macroscopically, the role of the system size is not fully understood and certainly needs clarification, also with regard to the morphological phase diagram. Furthermore, it is still an open question how other geometrical parameters such as the injection angle and cavity shape influence the morphological phase diagram.
\begin{acknowledgments}
This work was supported by the Grant TH-06 07-3 of the Swiss Federal Institute of Technology Zurich.
\end{acknowledgments}

\end{document}